\documentclass[twocolumn,english]{article}
\usepackage[T1]{fontenc}
\usepackage[latin9]{inputenc}
\usepackage{verbatim}
\usepackage{amstext}
\usepackage{graphicx}

\makeatletter
\newcommand{\lyxaddress}[1]{
\par {\raggedright #1
\vspace{1.4em}
\noindent\par}
}

\newcommand{\braket}[2]{\left \langle #1 \mid #2 \right \rangle}

\makeatother

\usepackage{babel}
\begin{document}

\title{Experimental Freezing of mid-Evolution Fluctuations with a Programmable
Annealer}

\author{Nicholas Chancellor{*}$^{\dagger\&}$, Gabriel Aeppli$^{\ddagger}$
and Paul A. Warburton$'{}^{\dagger}$}

\maketitle

\lyxaddress{$\dagger$London Centre For Nanotechnology 19 Gordon St, London UK
WC1H 0AH\\
 $^{\ddagger}$Department of Physics, ETH Zürich, CH-8093 Zürich,
Switzerland\\
 Department of Physics, École Polytechnique Fédérale de Lausanne (EPFL),
CH-1015 Lausanne, Switzerland\\
 Synchrotron and Nanotechnology Department, Paul Scherrer Institute,
CH-5232, Villigen, Switzerland\\
 'Department of Electronic and Electrical Engineering, UCL, Torrington
Place London UK WC1E 7JE\\
 $^{\&}$ Current Affiliation: Department of Physics, Durham University,
South Road, Durham, UK}
\begin{abstract}
For randomly selected couplers and fields, the D-Wave device typically
yields a highly Boltzmann like distribution \cite{Chancellor(2015)}
indicating equilibration. These equilibrated data however do not contain
much useful information about the dynamics which lead to equilibration.
To illuminate the dynamics, special Hamiltonians can be chosen which
contain large energy barriers \cite{Boixo2015,Boixo(2013)}. In this
paper we generalize this approach by considering a class of Hamiltonians
which map clusters of spin-like qubits (which we will henceforth refer
to as 'spins') into 'superspins', thereby creating an energy landscape
where local minima are separated by large energy barriers. These large
energy barriers allow us to observe signatures of the transverse field
frozen. To study these systems, we assume that the these frozen spins
are describes by the Kibble-Zurek mechanism \cite{Kibble1976} which
was originally developed to describe formation of topological defects
in the early universe. It was soon realized that it also has applications
in analogous superconductor systems \cite{Zurek1985,Kibble1985,Zurek1996,Zurek1993}
and later realized to also be important for the transverse field Ising
model \cite{Damski2005,Dziarmaga(2005)}. We demonstrate that these
barriers block equilibration and yield a non-trivial distribution
of qubit states in a regime where quantum effects are expected to
be strong, suggesting that these data should contain signatures of
whether the dynamics are fundamentally classical or quantum. We \emph{exhaustively}
study\emph{ }a class of 3x3 square lattice superspin Hamiltonians
and compare the experimental results with those obtained by exact
diagonalisation. We find that the best fit to the data occurs at finite
transverse field. We further demonstrate that under the right conditions,
the superspins can be relaxed to equilibrium, erasing the signature
of the transverse field. These results are interesting for a number
of reasons. They suggest a route to detect signatures of quantum mechanics
on the device on a statistical level, rather than by observing the
behavior for specially chosen Hamiltonians, as was done in \cite{Boixo2015,Lanting2014}.
Furthermore, our work suggests that devices of this kind may be able
to provide a way of studying the Kibble-Zurek mechanism in large and
complex systems, which may be interesting in its own right due to
the relevance of Kibble-Zurek to aspects of cosmology as well as condensed
matter physics. Finally the Ising square lattice with random fields
and couplers is known to be an NP-hard problem \cite{Barahona1982},
meaning that this class of Hamiltonians could provide a potential
avenue to study the effect of dynamical freezing on computation.
\end{abstract}

\section{Introduction\label{sec:Introduction}}

To understand the dynamics of a given system, be it the early universe,
or a condensed matter system, it is important to have data which showcase
non-equilibrium distributions. Equilibrated data only provide information
about the free energy landscape and provide a lower bound on the relaxation
rates (the system had to relax fast enough, otherwise one wouldn't
see equilibrium). Non-equilibrated data can be understood through
the Kibble-Zurek mechanism, in which relaxation rates are assumed
to change quickly from an 'adiabatic' regime, where relaxation rates
are fast compared to the relevant timescales of the evolution to an
'impulse' regime, where the dynamics are effectively frozen \cite{Damski2005}.
We call this transition the 'freezing' transition.

\begin{figure}
\begin{centering}
\includegraphics[width=7cm]{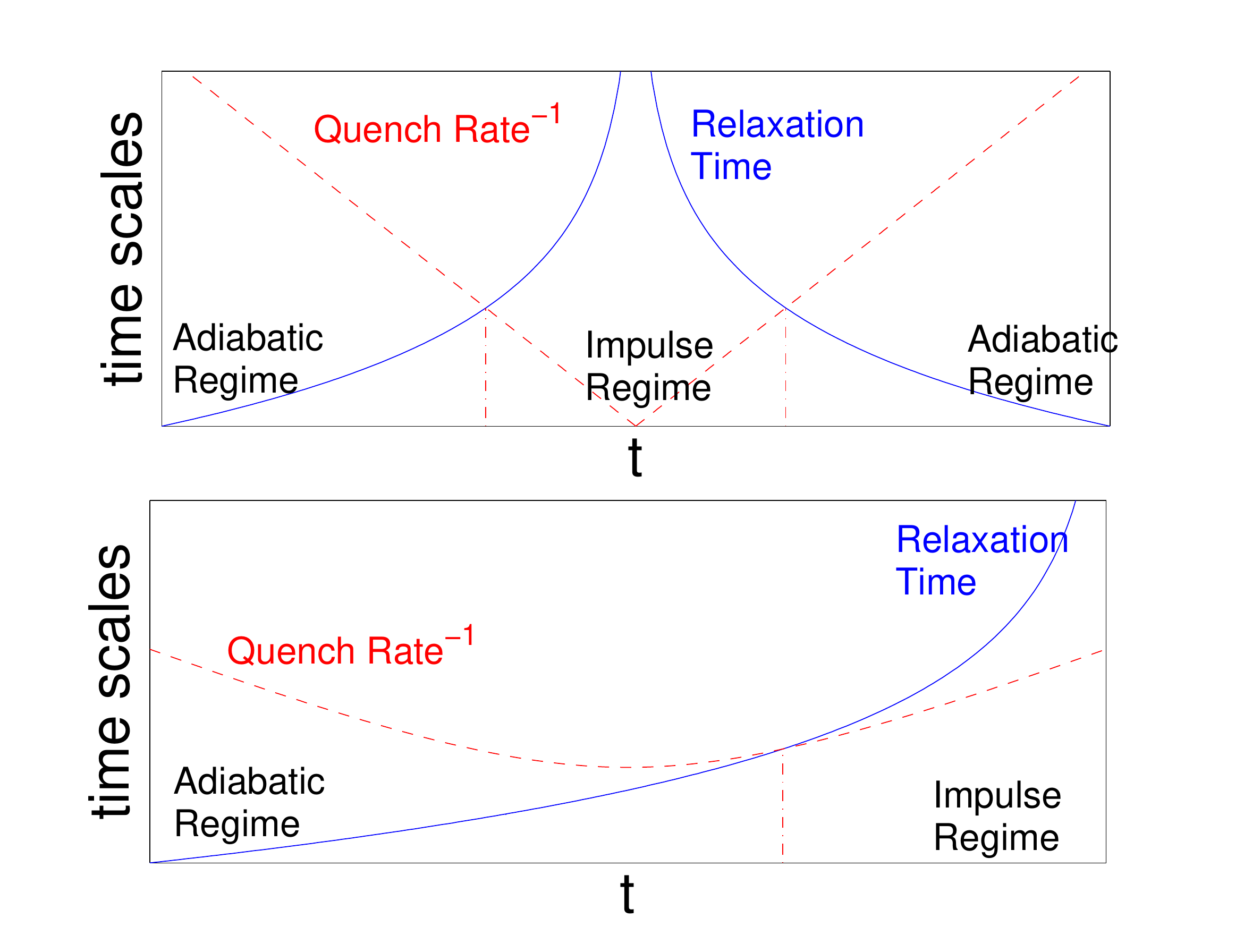}
\par\end{centering}

\caption{\label{fig:KZ_cartoons}Top, schematic representation of typical Kibble-Zurek
mechanism for defect generation: near a phase transition critical
slowing down causes the dynamics of a system to be effectively 'frozen'.
Crossing this frozen regime leads to topological defects within the
system. Bottom, schematic of Kibble-Zurek mechanism for annealer:
relaxation time increases as the transverse field is reduced, when
this timescale matches the inverse annealing rate, freezing occurs.
Frozen in spin configurations can be viewed as ``defects'' . }

\end{figure}

The Kibble-Zurek mechanism was first introduced in the context of
topological defect formation in the early universe \cite{Kibble1976}.
When a quenched system crosses a phase transition, the relaxation
time diverges. During this 'critical slowing down' a transition between
the 'adiabatic' and 'impulse' regime can be seen as occurring at a
freeze time when the relaxation timescale matches the rate of change
of the quench as illustrated in Fig. \ref{fig:KZ_cartoons}. Any topological
defects which are present when this transition occurs will be trapped,
even if the system enters the adiabatic regime on the other side of
the phase transition \cite{Zurek1996}. The density of topological
defects in the final system is therefore determined by the system
at the freeze time, which is in turn related to the rate of the quench.
For the annealer there is no re-entry into a new adiabatic regime
as there is with defect formation from crossing a phase transition.
In this simple case, we approximate the final spin configuration as
being the configuration at the freezing point. If freezing happens
at a point where quantum fluctuations are still strong, then traces
of these fluctuations should be visible at the end of the anneal.
As we shall discuss later, we can even define defects on the annealer
which behave in an analogous way to the topological defects considered
in a more traditional setting for the Kibble-Zurek mechanism.

It is appealing to use this celebrated mechanism to greatly simplify
the task of understanding the behavior of a complex system such as
the D-Wave device. To this end we consider the simplest possible incarnation
of this mechanism. We approximate that under a restricted set of circumstances,
to be discussed later, the transition from the 'adiabatic' to 'impulse'
regime occurs instantaneously, and always at the same effective temperature
and strength of transverse field. While not completely accurate, we
will demonstrate that this approximation does capture some important
features of the behavior of the device.

The experimental device we use is designed to implement a time dependent
transverse Ising model of the following form

\begin{equation}
H(t)=-A(t)\sum_{i}\sigma_{i}^{x}+\alpha B(t)H_{Ising}^{chi}\label{eq:Ht}
\end{equation}

\begin{equation}
H_{Ising}^{chi}=(\sum_{i}h_{i}^{chi}\sigma_{i}^{z}-\sum_{i,j\in\chi}J_{ij}^{chi}\sigma_{i}^{z}\sigma_{j}^{z})\label{eq:Hchi}
\end{equation}

\begin{figure}
\begin{centering}
\includegraphics[width=7cm]{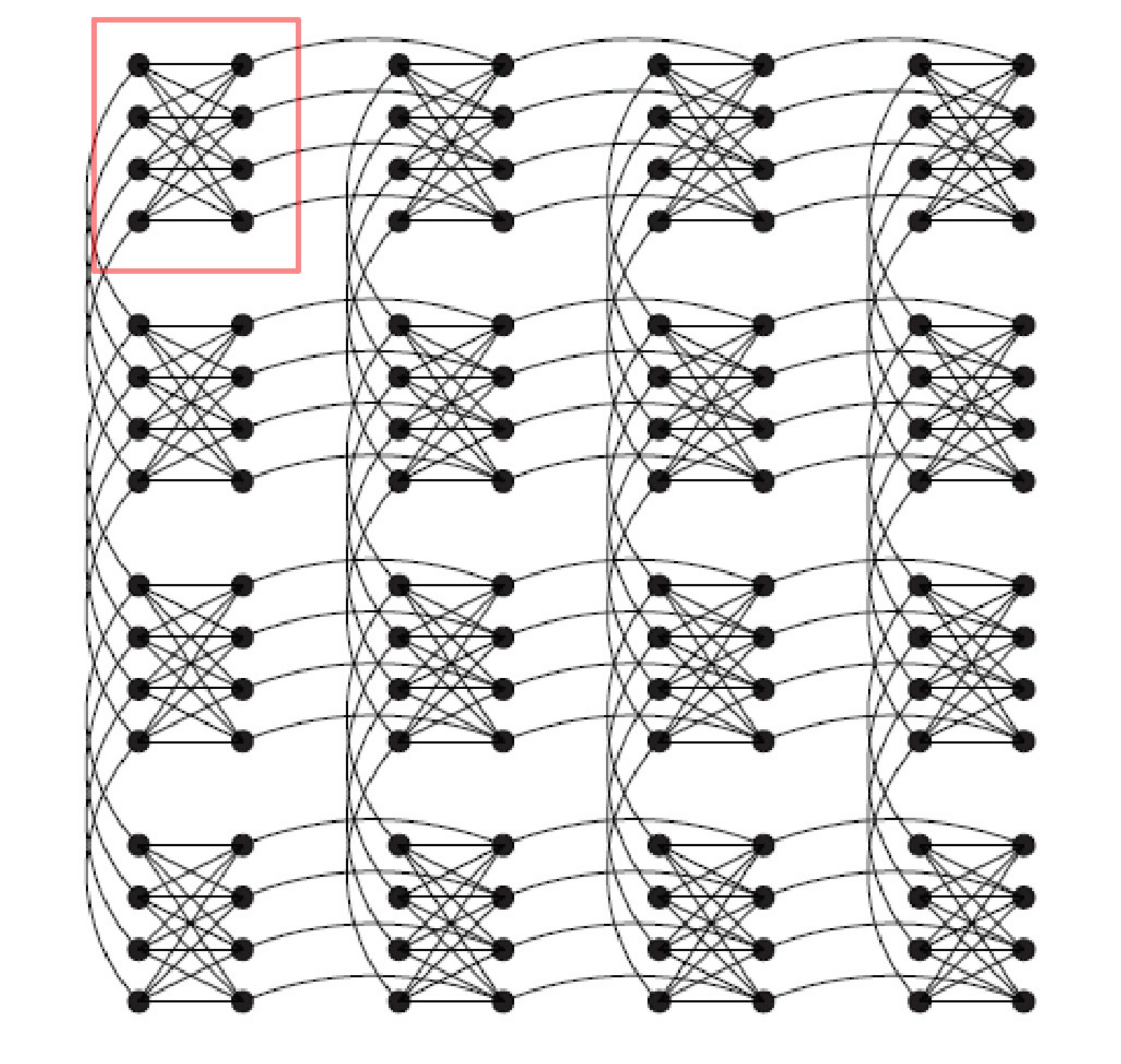} 
\par\end{centering}

\caption{\label{fig:chimera4x4}Chimera graphs used for this study. The full
figure shows a 4x4 array of unit cells each containing eight spins,
a single example of which is shown in the rectangle at the top left
of the figure. Each dot represents a spin and each line a coupler.}
\end{figure}

Here $\sigma^{x}$ and $\sigma^{z}$ are the Pauli spin matrices,
$J_{ij}^{chi}$ and $h_{i}^{chi}$ are user-programmable local couplers
and fields respectively. $\alpha$ sets the overall Ising energy scale.
The connectivity of the D-Wave machine is described by the so-called
Chimera graph, \textit{$\chi$, }as shown in Fig. \ref{fig:chimera4x4}\textit{.
}During the course of a single optimization run, \textit{A}(\textit{t})
is adiabatically reduced to near zero in a manner analogous to simulated
annealing in which the scale of thermal fluctuations (\textit{i.e.}
the temperature) is reduced to zero. At $t=t_{f}$ the system is described
by $H(t)\approx\alpha B(t)\,H_{Ising}^{chi}$ and the dynamics are
fully classical. Unfortunately, the rates at which $A(t)$ and $B(t)$
can be changed experimentally is limited, and therefore a 'typical'
case of randomly selected fields and couplers for $H_{Ising}^{chi}$
will yield an equilibrium result, even at the fastest allowed sweep
rate \cite{Chancellor(2015)}.

It has already been demonstrated however \cite{Boixo2015} that flipping
a fully connected chimera unit cell creates an energy barrier which
is high enough to prevent equilibration. Based on this observation,
we replace $H_{Ising}^{chi}$ in Eq. \ref{eq:Hchi} by a restricted
set of 'superspin' Hamiltonians $H_{Ising}^{ss}$:

\begin{equation}
H(t)=-A(t)\sum_{i}\sigma_{i}^{x}+\alpha B(t)\,H_{Ising}^{ss}.
\end{equation}
Here all couplers within a unit cell are of maximum allowed strength
and ferromagnetic 
\begin{equation}
H_{Ising}^{ss}=-\sum_{i,j\in\chi_{int}}\sigma_{i}^{z}\sigma_{j}^{z}+\alpha_{s}H_{ss},\label{eq:ISGham}
\end{equation}

and

\begin{equation}
H_{ss}=(\sum_{i}\frac{1}{2}\sigma_{i}^{z}-\sum_{i,j\in\chi_{ext}}J_{ij}\sigma_{i}^{z}\sigma_{j}^{z}).
\end{equation}

\begin{figure}
\begin{centering}
\includegraphics[width=7cm]{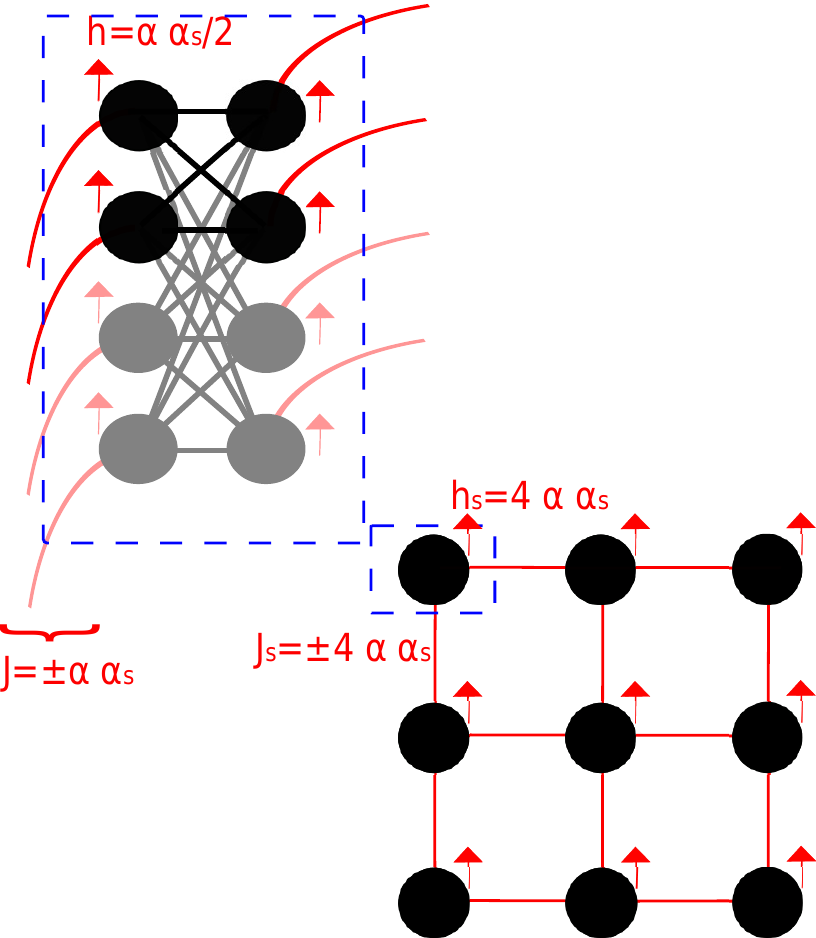}
\par\end{centering}

\caption{\label{fig:superspin mapping} Mapping of an effective square lattice
Hamiltonian onto the chimera graph using superspins. Runs with the
truncated unit cell utilize only the full colour part of the cell,
while runs with the whole cell include the faded part. All internal
couplers are ferromagnetic and of strength $\alpha$.}

\end{figure}

In this case we divide the graph $\chi\equiv\chi_{int}\cup\chi_{ext}$,
where $\chi_{int}$ indicate the couplers within a unit cell and $\chi_{ext}$
the couplers between unit cells. $\alpha_{s}\cdot\alpha$ sets the
overall energy scale of the superspin Hamiltonian. Hamiltonians of
this type will have an energy landscape in which every state for which
all of the spins within a unit cell agree is a local minimum. We can
therefore think of the low energy dynamics of this Hamiltonian as
that of an effective Hamiltonian in which each unit cell behaves like
a single spin. The underlying graph of these superspin Hamiltonians
is a square lattice, as demonstrated in Fig. \ref{fig:superspin mapping}. 

If the entire chimera graph is used in this way, then the overall
field for each superspin is $h_{s}=4\alpha\cdot\alpha_{s}$ while
the energy of each of the couplers is $J_{s}=\pm4\alpha\cdot\alpha_{s}$.
On the other hand if we consider truncated chimera unit cells consisting
of only 4 spins the fields and couplers take values $h_{s}=2\alpha\cdot\alpha_{s}$,
while the energy of each of the couplers is $J_{s}=\pm2\alpha\cdot\alpha_{s}$.
We choose $|h_{s}|=|J_{s}|$ so that the effective superspin Hamiltonian
is reminiscent of the Hamiltonian used in \cite{Chancellor(2015)}.
As was the case for the single chimera unit cell in\cite{Chancellor(2015)},
we can simplify our task by examining only symmetry inequivalent Hamiltonians.
For the 3x3 square lattice this use of symmetry equivalent reduces
the number of Hamiltonians we must investigate from $2^{12}$ to a
more manageable $570$. Because of the local gauge symmetry, this
is also the set of symmetry inequivalent Hamiltonians for the larger
($2^{21}$) set in which $h_{s}\propto\pm\alpha\cdot\alpha_{s}$.

\subsection*{Experimental methods}

All experiments were performed on the D-Wave Vesuvius processor located
at the Information Sciences Institute of the University of Southern
California. This processor contains 512 bits in an 8x8 array of unit
cells. Due to fabrication errors, nine of the bits fall outside of
the acceptable calibration range. We elected to study a 3x3 patch
of unit cells which did not contain any of these defective bits. The
annealing time $t_{f}$ was fixed at $20\mu s$ except where otherwise
stated. All data were sampled over randomized gauges and transformations
under the dihedral symmetry of the square lattice.

\section{Calculation Method: spin-sign transitions}

Motivated by the relationship to real world inference problems \cite{Chancellor(2015)}
and the fact that we desire to look at discrete rather than continuous
problems, we chose to restrict our analysis to examining the sign
of mean spin orientations. Examining the sign only rather than the
full mean orientation simplifies our analysis greatly, as knowledge
of the locations where the mean orientation of the spins pass though
zero plus the sign of the orientation at high temperature allow complete
knowledge of the sign of the orientation for the entire parameter
space. We refer to these zeros as \emph{spin-sign transitions}. 

\begin{figure}
\begin{centering}
\includegraphics[width=7cm]{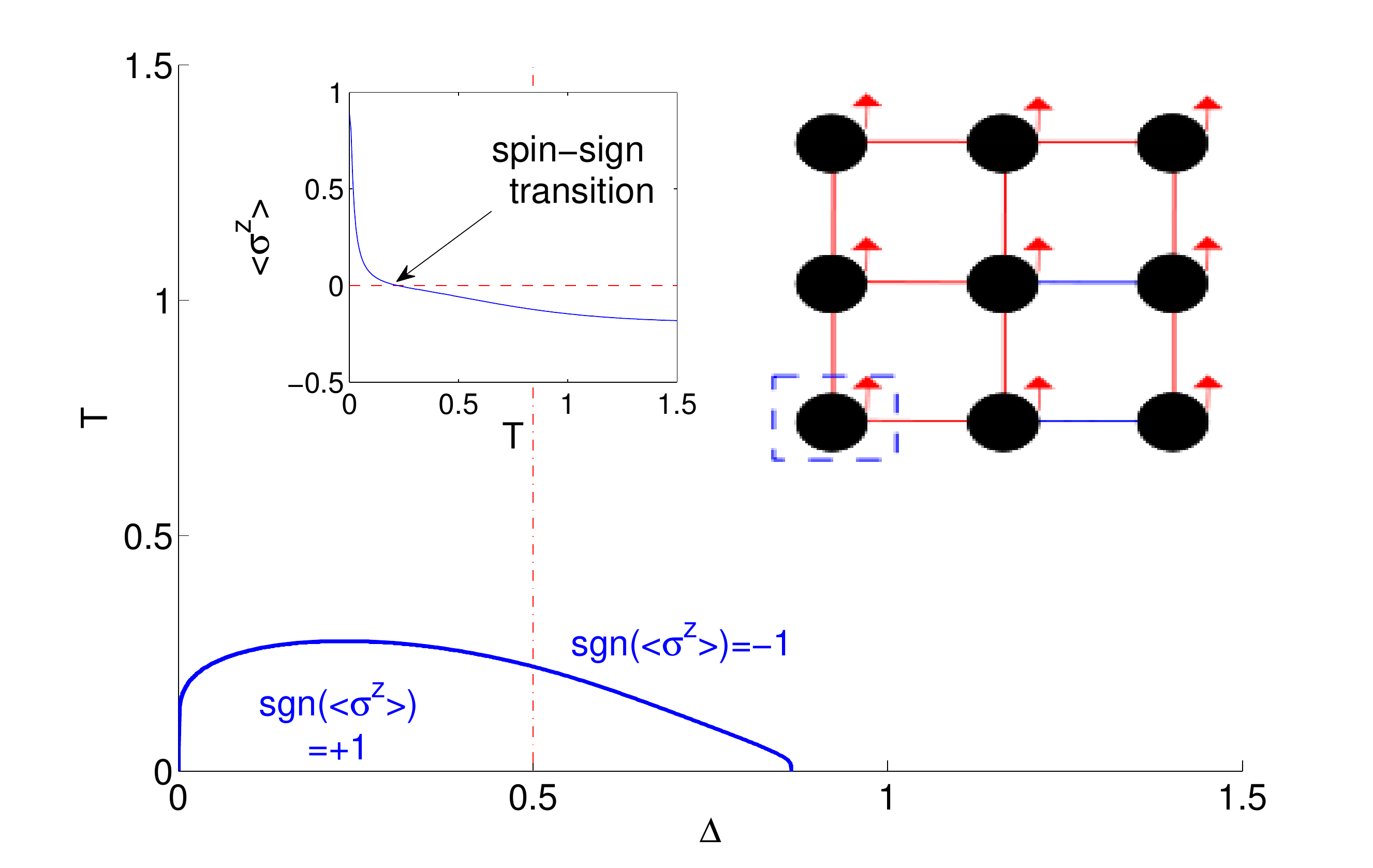}
\par\end{centering}

\caption{\label{fig:ss_example} An example of a spin-sign transition, which
divides regions of phase space where $\textrm{sgn}(\langle\sigma^{z}\rangle)=+1$
and $\textrm{sgn}(\langle\sigma^{z}\rangle)=-1$ calculated using
exact diagonalization. The transition occurs when the mean orientation
of the spin in the lower left corner of the 3x3 Hamiltonian shown
in the inset changes sign. All fields and couplings have the same
magnitude, with red indicating anti-ferromagnetic coupling and blue
indicating ferromagnetic. The inset plot shows the polarization versus
temperature on the $\Delta=0.5$ line and the associated spin-sign
transition.}

\end{figure}

Each spin within a Hamiltonian will have a finite (possibly empty)
set of spin-sign transitions with temperature. This concept can be
easily extended to include transverse field, with spin sign-transitions
as 1 dimensional curves in $T$-$\Delta$ space, rather than points.
In one dimension the spin-sign transitions are the nodes of a function,
in 2D the nodes become 1 dimensional lines. For the effective superspin
Hamiltonian shown in Fig. \ref{fig:superspin mapping}, these transitions
can be readily calculated using exact diagonalization. These spin-sign
transitions (along with the sign of the orientation somewhere not
directly on a transition) allow us to know the sign of the orientation
of every spin for any value of temperature and transverse field. Fig.
\ref{fig:ss_example} shows an example of a single spin-sign transition,
appearing as a line which divides regions of phase space where $\textrm{sgn}(\langle\sigma^{z}\rangle)=+1$
and $\textrm{sgn}(\langle\sigma^{z}\rangle)=-1$.

\begin{figure}
\begin{centering}
\includegraphics[width=7cm]{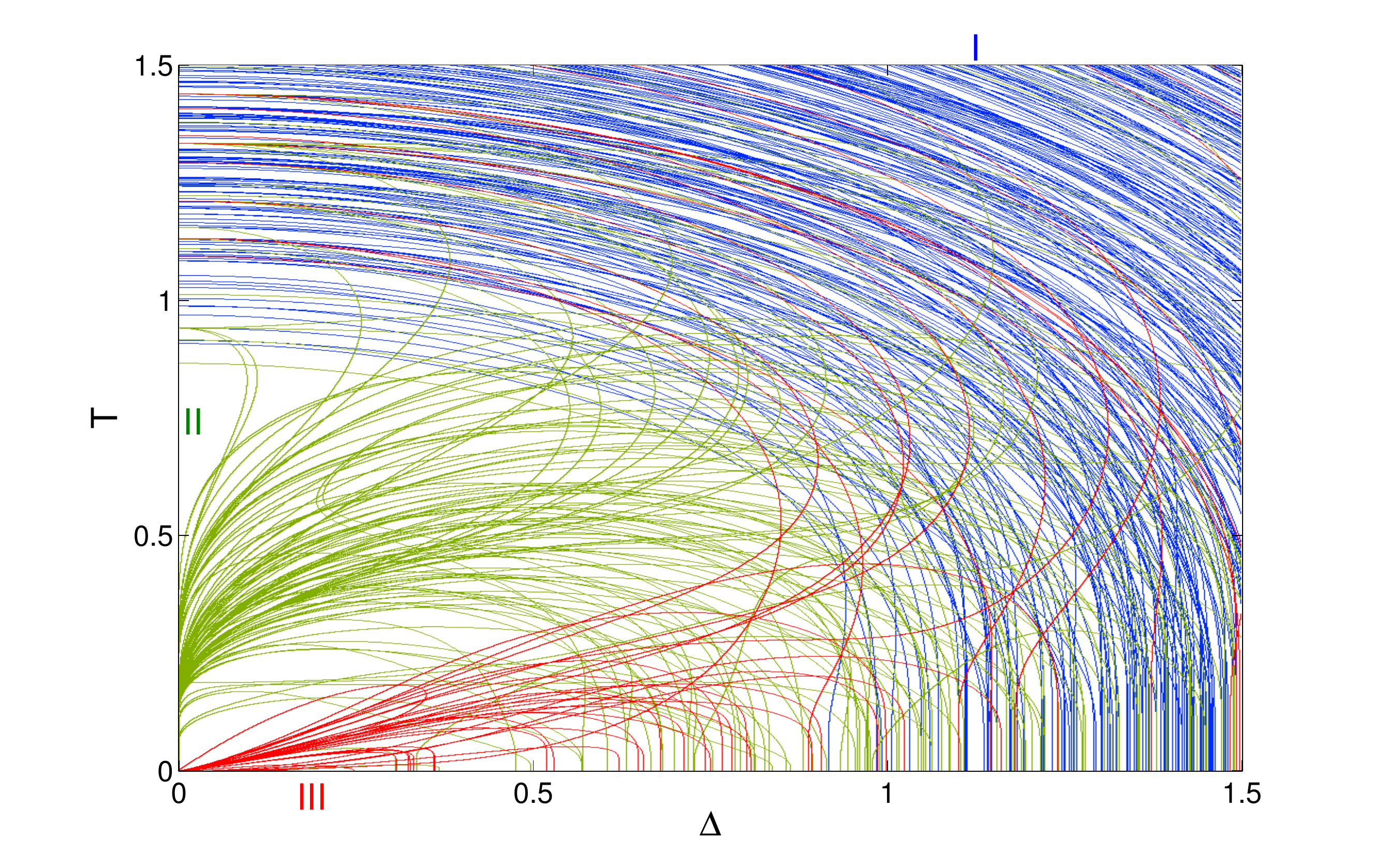}
\par\end{centering}

\caption{\label{fig:trans_dens} Spin-sign transitions for a 3x3 transverse
field Ising square lattice, exhaustively calculated for all $570$
symmetry inequivalent Hamiltonians. Colours indicate whether transitions
are associated with type I II or III spins.}

\end{figure}

We have two complementary goals for our spin-sign transition analysis,
firstly we want to be able to define defects for the Kibble-Zurek
mechanism to act upon and estimate the 'freeze time' when the dynamics
become slow enough that the system can be viewed as 'frozen'. We define
a \emph{spin-sign defect} as a spin for which the orientation has
a different sign then the one predicted for the final (finite temperature)
configuration at the end of the anneal. If the system remained perfectly
adiabatic throughout the evolution, then there would be no spin-sign
defects, analogous to topological defects in a more traditional setting
for the Kibble-Zurek mechanism. This analysis assume simple monotonic
behavior of the defect density, and therefore is most appropriate
for spins whose spin-sign transitions form simple arc like paths in
$T$-$\Delta$ space, as the transitions labeled I in Fig. \ref{fig:trans_dens}
do. We term this type of analysis 'defect rate analysis'. 

On the other hand, we want to be able to compare our experimental
results to static exact diagonalization calculations at a range of
temperature and transverse field values and be able to detect signatures
of the conditions when the system froze. One concern, especially in
light of the analysis performed in \cite{Otsubo(2012),Otsubo(2014)}
is that there will be little difference between the effect of finite
temperature and finite transverse field, and spin-sign transitions
will not provide a reliable way of distinguishing between thermal
and transverse field effects. Fig. \ref{fig:trans_dens} demonstrates
that the spins labeled II and\emph{ }III\emph{ }have complex structures
in their spin-sign transitions which will allow us to distinguish
these two effects. We term this type of analysis 'frozen spin analysis'.

Many of the spin-sign transitions however do form arc-like paths in
$T$-$\Delta$ space consistent with the results from \cite{Otsubo(2012),Otsubo(2014)}
and that for higher $T$ and $\Delta$ these transitions dominate
completely, see Sec. 1 of the supplemental material. 

Before going further, we need to discuss the method which we have
employed in classifying these spins. We first note that the highly
non-arc-like spin-sign transitions tend to cross through the origin
point, ($T,\Delta=0$). It is therefore interesting to consider these
spins separately from those which do not have a transition which crosses
the origin. A spin will have a spin-sign transition at the origin
if it does not have a definite \emph{ground state orientation;} in
other words, spins for which there are an equal number of states with
spin up as down in the zero transverse field ground state manifold.
However, this is not the only way in which that feature can come about.
If the point $T,\Delta=0$ is approached classically, i.e. by reducing
temperature at zero transverse field, the orientation will agree with
the ground state orientation. On the other hand approaching this point
quantum mechanically, by reducing transverse field at zero temperature,
holds no such guarantee. The ground state orientation at infinitesimal
transverse field therefore need not have the same sign of an unweighted
sampling of the classical ground state manifold, leading to a spin
sign transition at zero temperature and infinitesimal $\Delta$. We
therefore divide the spins into 4 types:
\begin{description}
\item [{$\textrm{T}\textrm{ype}\:\textrm{0}$}] No spin-sign transitions,
the sign of the orientation does not change at any value of $T$ or
$\Delta$. These spins are of no interest to this study. Of all $5130$
spins considered $3835$ have no spin-sign transitions between $0\leq T<5$
and $0\leq\Delta<5$, and therefore can be considered to be type 0
for our purposes.
\item [{$\textrm{T}\textrm{ype}\:\textrm{I}$}] Spin-sign transitions present,
but no spin-sign transition at $T,\Delta=0$. These transitions form
arc-like paths far from the origin, consistent with the behaviour
seen in \cite{Otsubo(2012),Otsubo(2014)}. We found $844$ spins for
which there were spin-sign transitions in the range $0\leq T<5$ and
$0\leq\Delta<5$ but which did not have a transition at $T,\Delta=0$.
These are the spins of interests for defect rate analysis.
\item [{$\textrm{T}\textrm{ype}\:\textrm{II}$}] No definite ground state
orientation, there are an equal number of states with the spin up
and spin down in the zero transverse field ground state manifold,
by construction these have a spin-sign transition at ($T,\Delta=0$).
We observe that $411$ of the spins are type II. From Fig. \ref{fig:trans_dens}
we can see that these spins possess a complex spin-sign transition
structure which is not desirable for defect rate analysis, but is
highly desirable in frozen spin analysis.
\item [{$\textrm{T}\textrm{ype}\:\textrm{III}$}] Definite ground state
orientation, but quantum fluctuations from an infinitesimal $\Delta$
stabilize an orientation with the opposite sign. We observe that $40$
of the spins are type III. From Fig. \ref{fig:trans_dens} we can
see that these spins also possess a complex spin-sign transition structure
which is not desirable for defect rate analysis, but is highly desirable
in frozen spin analysis.
\end{description}
There is no restriction the spin-sign transitions for a single need
to form a connected graph, therefore spins of type II and III may
still also have arc-like transitions far from the origin.

For the defect rate analysis we only want to count defects for type
I spins. For the frozen spin analysis on the other hand, we compare
the sign of the experimentally observed orientation of all type I,
II and III superspins to the theoretical equilibrium value from the
effective superspin Hamiltonian at different values of temperature
and transverse field. The spin orientations of these Hamiltonians
can be readily calculated by exact diagonalization. It is worth emphasizing
that this is a completely static quantity, and while there is theoretical
justification for comparing this quantity to our experimental data,
such comparison represents a gross simplification.

A highly simplified version of Kibble-Zurek where all Hamiltonians
with the same $\alpha$ and $\alpha_{s}$ all transition instantly
from the 'adiabatic' to the 'impulse' regime, (i .e. freeze) at the
same point in the annealing process predicts perfect agreement at
some value of $T$ and $\Delta$. It is in this sense that the frozen
spin analysis analysis assumes a very simple version of Kibble-Zurek.
In Sec. \ref{sec:Frozen-in-Behavior} we will show data which suggests
that, while not completely accurate, this highly simplified picture
does allow us to obtain non-trivial signatures of the behavior of
the spins at the point when they freeze. However before this we perform
a more traditional Kibble-Zurek analysis in Sec. \ref{sec:Rate-of-Spin-Sign}.

\section{Rate of Spin-Sign defects\label{sec:Rate-of-Spin-Sign}}

Before we compare in detail the spin orientations at every point in
$T$-$\Delta$ space, it is worth considering a defect density calculation
in analogy to what is usually considered for a condensed matter or
cosmological system undergoing the Kibble-Zurek mechanism. To do this
we must define a defect. We define a spin-sign defect as a spin for
which the orientation has a different sign then the one predicted
for the final (finite temperature) configuration at the end of the
anneal. If the system remained perfectly adiabatic throughout the
evolution, then there would be no spin-sign defects, analogous to
topological defects in a more traditional setting for the Kibble-Zurek
mechanism. 

\begin{figure}
\begin{centering}
\includegraphics[width=7cm]{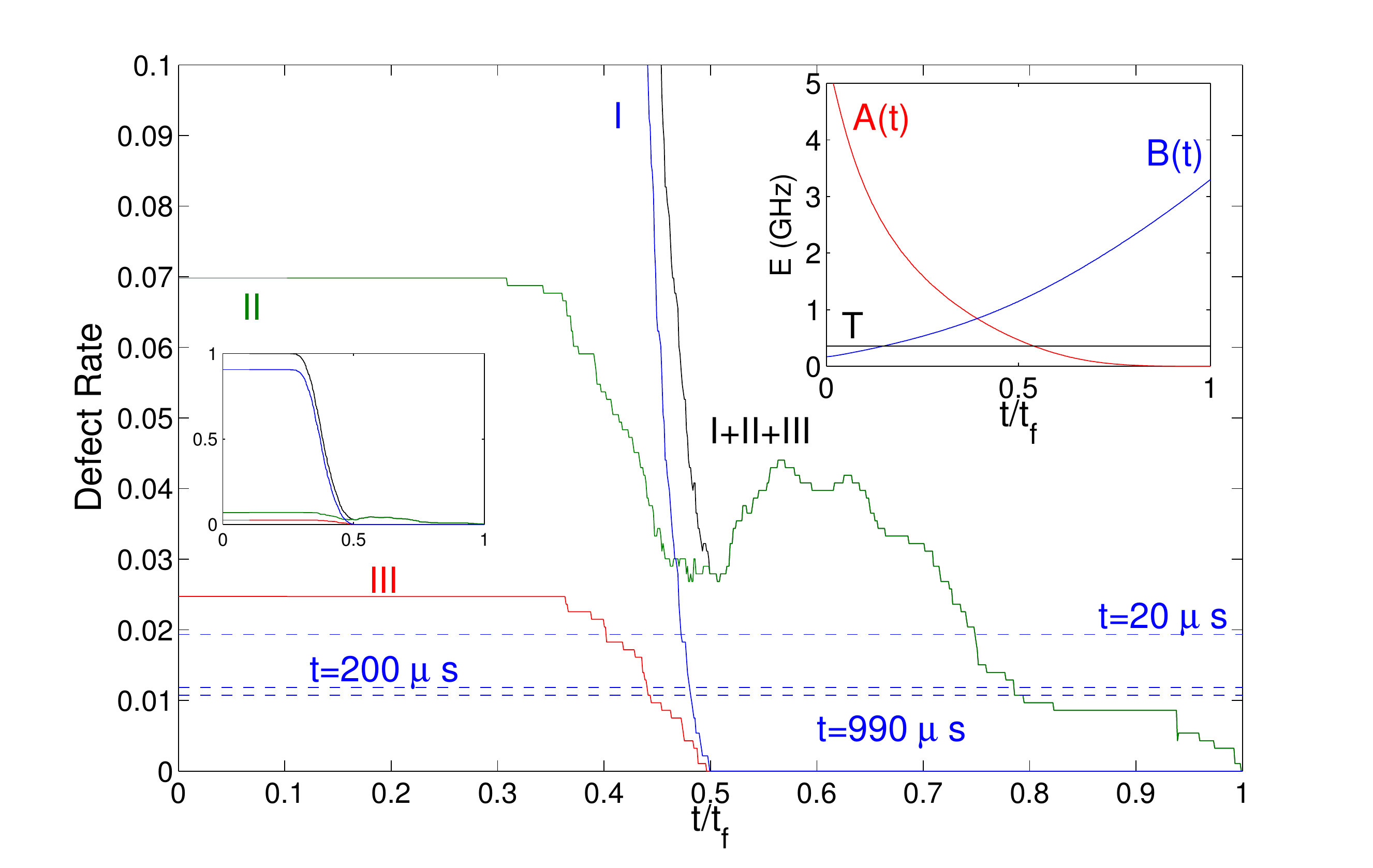}
\par\end{centering}

\caption{\label{fig:defect_ratec2}Equilibrium defect rate for different spin
types in the effective (9 spin) superspin Hamiltonian for $\alpha=0.25$,
$\alpha_{s}=1$ with a truncated 4-spin unit cell versus $t/t_{f}$.
A rate of $1$ corresponds to defects on $931$of the spins. Dotted
lines are the number of defects in type I spins observed in experimental
runs over $20\mu s$, $200\mu s$, and $990\mu s$. The number of
defects decreases with slower anneal time, as predicted by the Kibble-Zurek
mechanism. The black line represents all defects, while the blue represents
type I only, the gold type II and the red type III. Left inset: full
range of y axis. Right inset: annealing schedule.}

\end{figure}

One complication with defining spin-sign defects in such a way is
that these defects will not necessarily decrease monotonically in
number throughout the annealing process. Fortunately, we observe that
type I spins have highly arc-like spin-sign transitions, and therefore
should show a monotonic decrease in equilibrium defect number during
the annealing protocol, Fig. \ref{fig:defect_ratec2} shows that this
is indeed the case. The number of defects on type I spins decreases
with increasing annealing time, consistent with the Kibble-Zurek mechanism,
and suggests an effective 'freeze time' for (at least a subset of)
type I spins around $t/t_{f}=0.5$ which becomes slightly later as
$t_{f}$ is increased. Given the low rate of defects, another possible
interpretation is that freezing actually occurs at $t/t_{f}\gg0.5$,
and the defects seen here are anomalies which are not captured in
our simplified Kibble-Zurek picture. To answer determine which interpretation
is correct, we must perform a numerical estimate of the freeze time.

To perform a numerical estimate of the freeze time we ideally would
like to exactly diagonalize the 3x3 truncated chimera graph and use
open quantum system techniques to estimate the decay rate. Unfortunately,
the number of spins is outside of the range which we can numerically
access with standard ED techniques. To accurately describe the freezing
process however we want to go beyond the simple spin $\frac{1}{2}$
description of the superspins used elsewhere in this manuscript. 

To capture the internal freezing behavior of the superspins, while
still operating in a numerically tractable regime, we approximate
each truncated chimera unit cells as a $K(4)$ fully connected graph
in which each spin couples equally to the all of the spins in adjacent
superspins. We take the internal coupling strength to be the average
couplings between the spins in the chimera. This approximation endows
each unit cell with full permutation symmetry, meaning that as long
as the symmetry is unbroken each can be described by 5 states. This
is in contrast to the actual $K(2,2)$ graph of the truncated chimera
unit cell, for which 9 states are required. 

To estimate the freeze time we choose to focus on the Hamiltonian
for which all couplings are ferromagnetic, this allows us to further
reduce the size of the computation by taking advantage of the fact
that this choice of Hamiltonian does not break the dihedral symmetry
of the 3x3 square graph. For the relaxation timescale we calculate
the decay rate from the first excited state to the ground state using
standard Redfield formalism with realistic coupling parameters \cite{Amin_privComm,Breuer2002}
assuming a dominant coupling to the bath in the $\sigma^{z}$ direction. 

We approximate the quench rate as the magnitude of the overlap between
the ground and first excited state at times which are separated by
an infinitesimal time step $\Delta t$ divided by $\Delta t$,

\begin{equation}
\frac{|\braket{\psi_{0}(t)}{\psi_{1}(t+\Delta t)}|}{\Delta t}.
\end{equation}

\begin{figure}
\begin{centering}
\includegraphics[width=7cm]{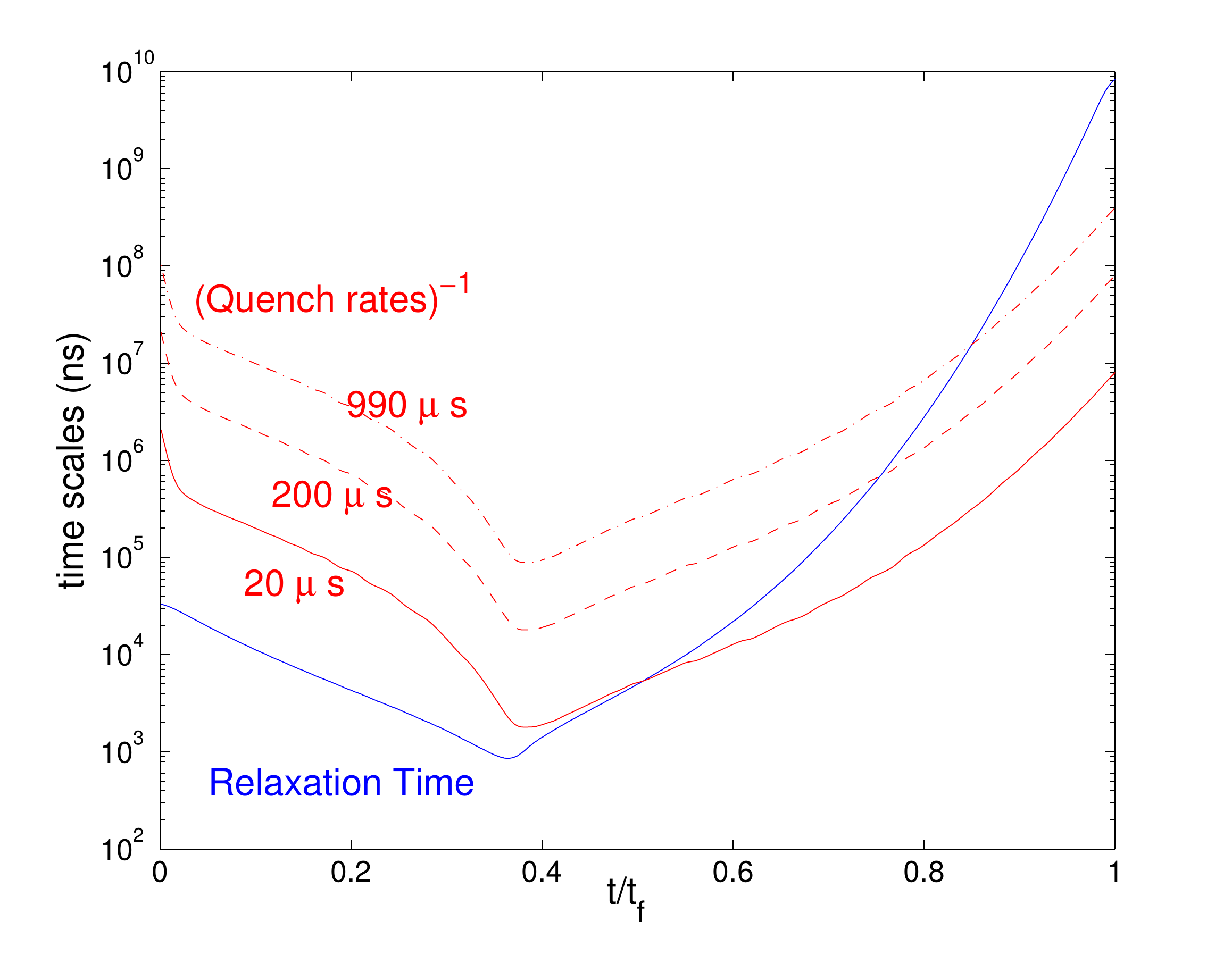}
\par\end{centering}

\caption{\label{fig:time_compare}Comparison of inverse quench rates approximated
by $\frac{|\braket{\psi_{0}(t)}{\psi_{1}(t+\Delta t)}|}{\Delta t}$
where $\Delta t=t_{f}/1000$ for the 3 annealing times examined in
Fig. \ref{fig:defect_ratec2} versus relaxation time induced by the
bath. This plot is based on an approximation where chimera unit cells
are treated as fully connected graphs. The Kibble-Zurek approximation
is to assume that the system freezes when the inverse quench rate
becomes equal to the relaxation time.}
\end{figure}

The numerical data in Fig. \ref{fig:time_compare} suggest that the
freeze time for $t_{f}=20\mu s$ is indeed around $t/t_{f}=0.5$,
and therefore the larger number of defects seen for this anneal time
(compared with $t_{f}=200\mu s$ and $t_{f}=990\mu s$) can be explained
by the Kibble-Zurek mechanism assuming all Hamiltonians freeze at
the same time. On the other hand, for $t_{f}=200\mu s$ and $t_{f}=990\mu s$,
the freeze time is predicted to be much later, indicating that the
type I defects seen in these cases are anomalies which are not well
described by one or more of the approximations we apply.

\section{Frozen in Behavior of Spins \label{sec:Frozen-in-Behavior}}

\begin{figure}
\begin{centering}
\includegraphics[width=7cm]{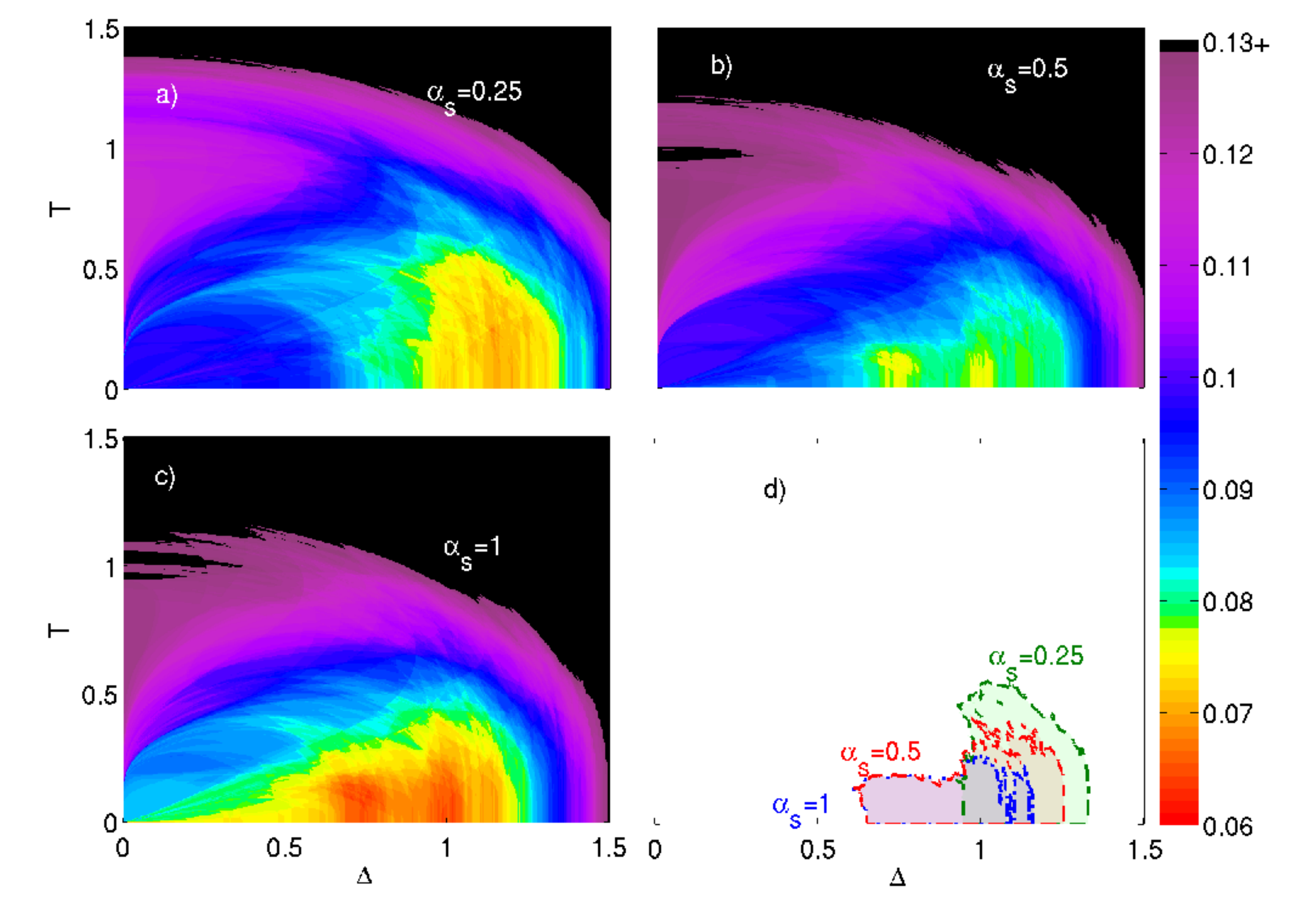}
\par\end{centering}

\caption{\label{fig:T_Delta_a_s} a-c) Colour plots indicating the fraction
of bits which disagree with finite temperature and transverse field
states for the effective superspin Hamiltonian for $\alpha=1$ and
different values of $\alpha_{s}$ as follows a) $\alpha_{s}=0.25$
b) $\alpha_{s}=0.5$ c) $\alpha_{s}=1$. $\Delta$ and $T$ are both
in dimensionless units. d) lines and shading indicate parameter regimes
where the number of disagreements is 10 or less away from the best
agreement. All data in this plot are for an anneal time of $20\,\mu s$.
Note that bits with no spin-sign transition were excluded from these
plots. $\Delta$ and $T$ are both in dimensionless units.}

\end{figure}

\begin{figure}
\begin{centering}
\includegraphics[width=7cm]{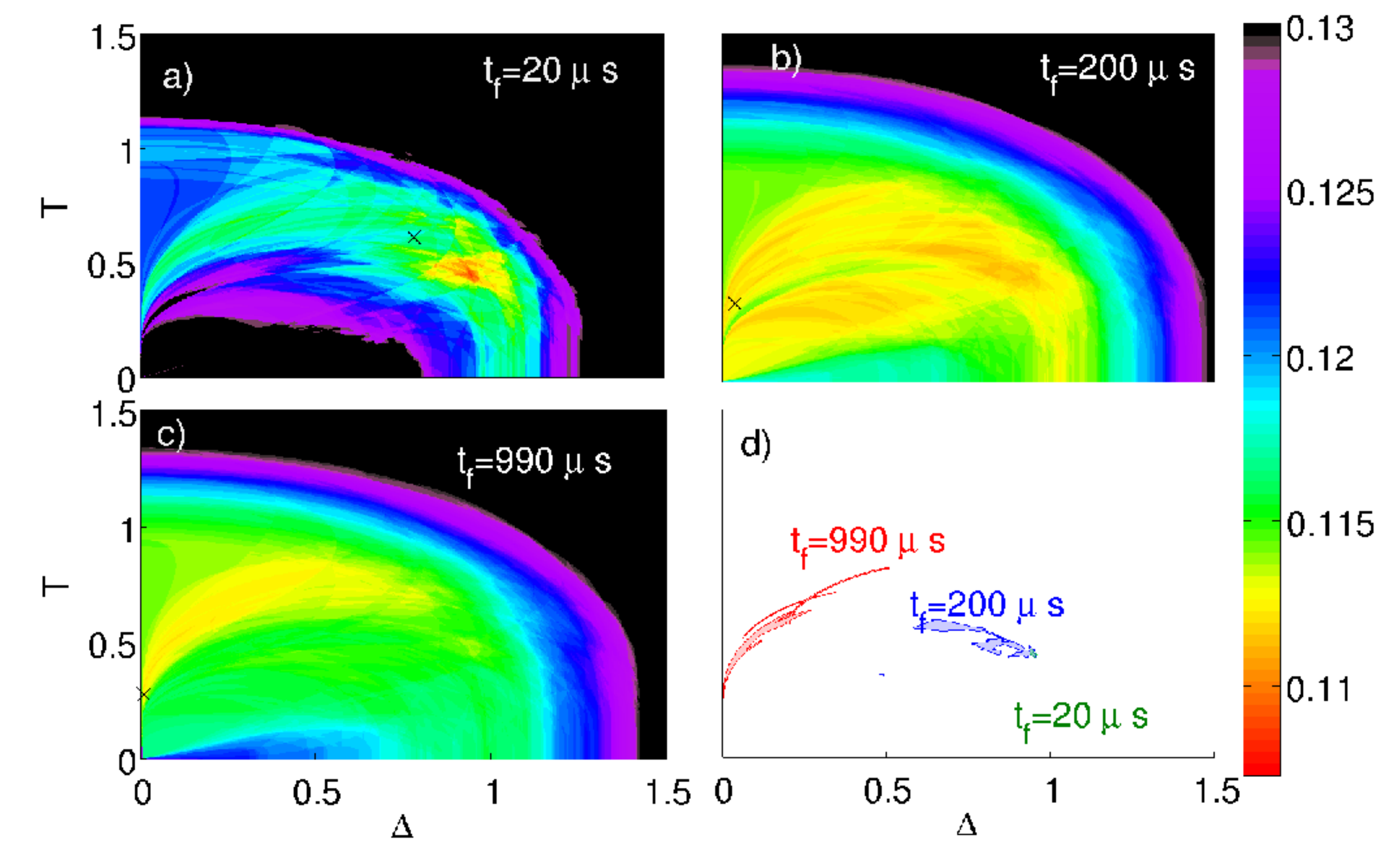}
\par\end{centering}

\caption{\label{fig:Delta_t_c2}Plots of truncated cell with two spins on each
side $\alpha=0.25$, $\alpha_{s}=1$ run for various anneal times.
a-c) Colour plots indicating the fraction of bits with spin sign transitions
which disagree with finite temperature and transverse field states
for the effective superspin Hamiltonian as follows b) $t=20\,\mu s$
c) $t=200\,\mu s$ d) $t=990\,\mu s$. d) lines and shading indicate
parameter regimes where the number of disagreements is 3 or less away
from the best agreement. $\Delta$ and $T$ are both in dimensionless
units. X indicates freezing parameters from Fig. \ref{fig:time_compare}.}
\end{figure}

Comparison between the experimental data and the static exact diagonalization
model appear in Figs. \ref{fig:T_Delta_a_s} and \ref{fig:Delta_t_c2}.
In these figures, the experimentally observed sign of the spin orientations
are compared with the static superspin model at various values of
temperature and transverse field. In subfigures b)-d) of these figures
the number of disagreements with the exact diagonalization model is
represented as a fraction of the $1295$ spins for which we observed
a spin-sign transition in the superspin model for $0\leq T<5$ and
$0\leq\Delta<5$. It is worth noting that we do not consider the $3835$
remaining (type 0) spins for which there is no transition, as the
sign of the orientation of these spins contains no useful information
as to the values of temperature and transverse field which the device
has experienced. Subfigure a) of these two figures plots the region
where the disagreement for the remaining subfigure is close to minimal,
and therefore facilitates direct comparison between the data in the
other subfigures.

Fig. \ref{fig:T_Delta_a_s} demonstrates that for $\alpha=1$ the
data agree best with the behavior at low temperatures and moderate
transverse field. Moreover, as $\alpha_{s}$ is reduced, the transverse
field for best agreement increases. This trend is what we should expect
in a Kibble-Zurek mechanism picture, where the energy scale of the
couplers in the superspin Hamiltonian $\alpha_{s}$ is reduced relative
to the transverse field. This indicates that our approach is successful,
on average, at trapping remnants of the transverse fields, and that
we have successfully captured data outside of the Boltzmann like equilibrium
expected at $t=t_{f}$. It is encouraging that the best agreement
is in a regime where transverse field is similar in strength to the
couplers, and temperature is low, as this is the regime where quantum
fluctuations are expected to be the strongest. While suggestive, this
fact in itself is not conclusive evidence of quantum behavior, as
there may be mean field models of the form \cite{Crowley2015,Shin(2014)}
which also behave in this way. Showing that we can access this regime
however is an interesting result in itself, as it suggests that further
analysis of this data will likely prove fruitful for examining the
role which quantum effects play.\footnote{Authors Note: We are currently in the process of obtaining access
to computing resources so that we can perform classical simulations
for comparison, results to appear in subsequent pre-prints and published
version.} Compared to the quantum calculation, the classical spin calculation
is much more numerically intensive, and we are currently working on
analysis in this direction.

These data are also interesting in that they show that a much simplified
Kibble-Zurek description of the device allows us to detect traces
of the dynamics in a non-trivial way. Unsurprisingly, however there
is no value of $T$ and $\Delta$ for which the data agree perfectly
with static conditions of finite temperature and transverse field.
This is to be expected given the large number of approximations which
must be made to assume that the chip data would agree perfectly. Here
we list some possible reasons for imperfect agreement: the internal
degrees of freedom of the superspins (the individual spins) may lead
to non-trivial corrections; not all Hamiltonians may freeze at the
same time; the superspins themselves may not freeze uniformly for
a given Hamiltonian; control errors may cause order-by-disorder effects
which do not average out and/or; Interaction with the bath may change
the Hamiltonian spectrum via the Lamb shift \cite{Albash2012}. 

To demonstrate that this effect is in fact Kibble-Zurek, we examine
the effect of longer annealing times. In Fig. \ref{fig:Delta_t_c2},
the data tend to agree more strongly with model with lower transverse
field as this time is increased. This is consistent with the data
having time to equilibrate during the longer anneals, and traces of
the transverse field being erased. 

Note that Fig. \ref{fig:Delta_t_c2} shows data with a truncated 4
spin unit cell and $\alpha=0.25$, $\alpha_{s}=1$, rather than the
full 8 spins. For the full 8 spin system, even at low values of $\alpha$,
equilibration could be observed for anneal times up to $2000\,\mu s$
(see Sec. 2 of the supplemental material). We suspect that the reason
for this is that we are unable to experimentally access the timescales
which are required to reach equilibrium. This is reasonable, because
for an isolated superspin (without fields) based on the full unit
cell flipping the superspin only comes in at fourth order of perturbation
theory and has an effective potential barrier of $16\,\alpha\,B(t)$.
Flipping the truncated version of the other hand only requires second
order in perturbation theory and comes with an energy barrier of only
$4\,\alpha\,B(t)$. 

\section{Conclusions}

We have demonstrated that we are able to detect remnants of the dynamics
in a non-trivial way by using a class of superspin Hamiltonians which
feature strong energy barriers. These data suggest that we are able
to preserve information from a regime of low temperature and moderate
transverse field, which is exactly the regime where quantum behaviors
are expected to be the strongest. With further numerical classical
spin calculations, one should be able to distinguish between classical
and quantum behavior. Moreover the observation of this mechanism on
the D-Wave device suggest that the device may be useful in studying
the Kibble-Zurek mechanism in highly complex Ising systems, an interesting
prospect in its own right, given the the importance of this mechanism
to aspects of cosmology and condensed matter physics. These experiments
are also tantalizing because the underlying superspin Hamiltonians
map to a class of NP-hard problems \cite{Barahona1982}, and therefore
may also be interesting in illuminating the effect which the Kibble-Zurek
mechanism may have on computation.

\subsection*{Acknowledgements}

This work was supported by Lockheed Martin and by EPSRC (grant refs:
EP/K004506/1 and EP/H005544/1). We thank the USC Lockheed Martin Quantum
Computing Center at the University of Southern California's Information
Sciences Institute for access to their D-Wave Two machine. We acknowledge
fruitful discussions with Andrew Green, Szilard Szoke, and Markus
Mueller and thank Mohammad Amin for providing the Redfield code on
which we based our relaxation time calculation.

\title{Supplemental Material for: Experimental Freezing of mid-Evolution
Fluctuations with a Programmable Annealer}

\section{Spin-sign transitions at higher temperature and transverse field}

Fig. 3 of the main manuscript illustrates the spin-sign transitions
for $0<T<1.5$ and $0<\Delta<1.5$. Fig. S1 illustrates that at higher
values of $T$ and $\Delta$, the transition densities follow arcs.
This is consistent with the results found in \cite{Otsubo(2012),Otsubo(2014)},
which predict little difference between the ability to decode using
thermal fluctuations and qunatum fluctuations.

\begin{figure}
\begin{centering}
\includegraphics[width=7cm]{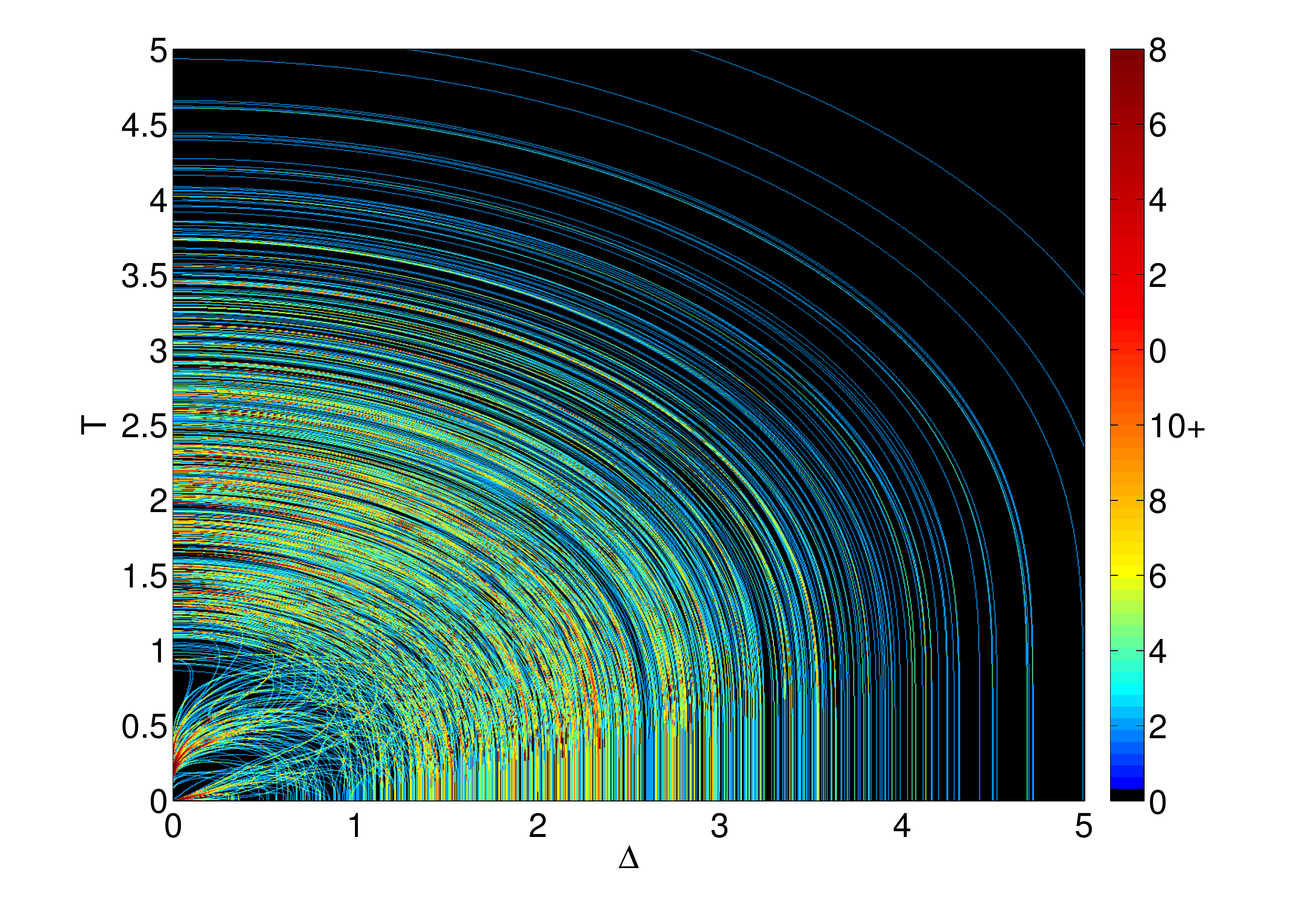}
\par\end{centering}

Figure S1: Density of spin sign transitions for a wider range of $T$
and $\Delta$ then Fig. 3 of the main text.
\end{figure}

Based on the shape of these transitions, we are not able to reliably
distinguish between the effects of temperature and transverse field,
as Fig. S2 demonstrates.

\begin{figure}
\begin{centering}
\includegraphics[width=7cm]{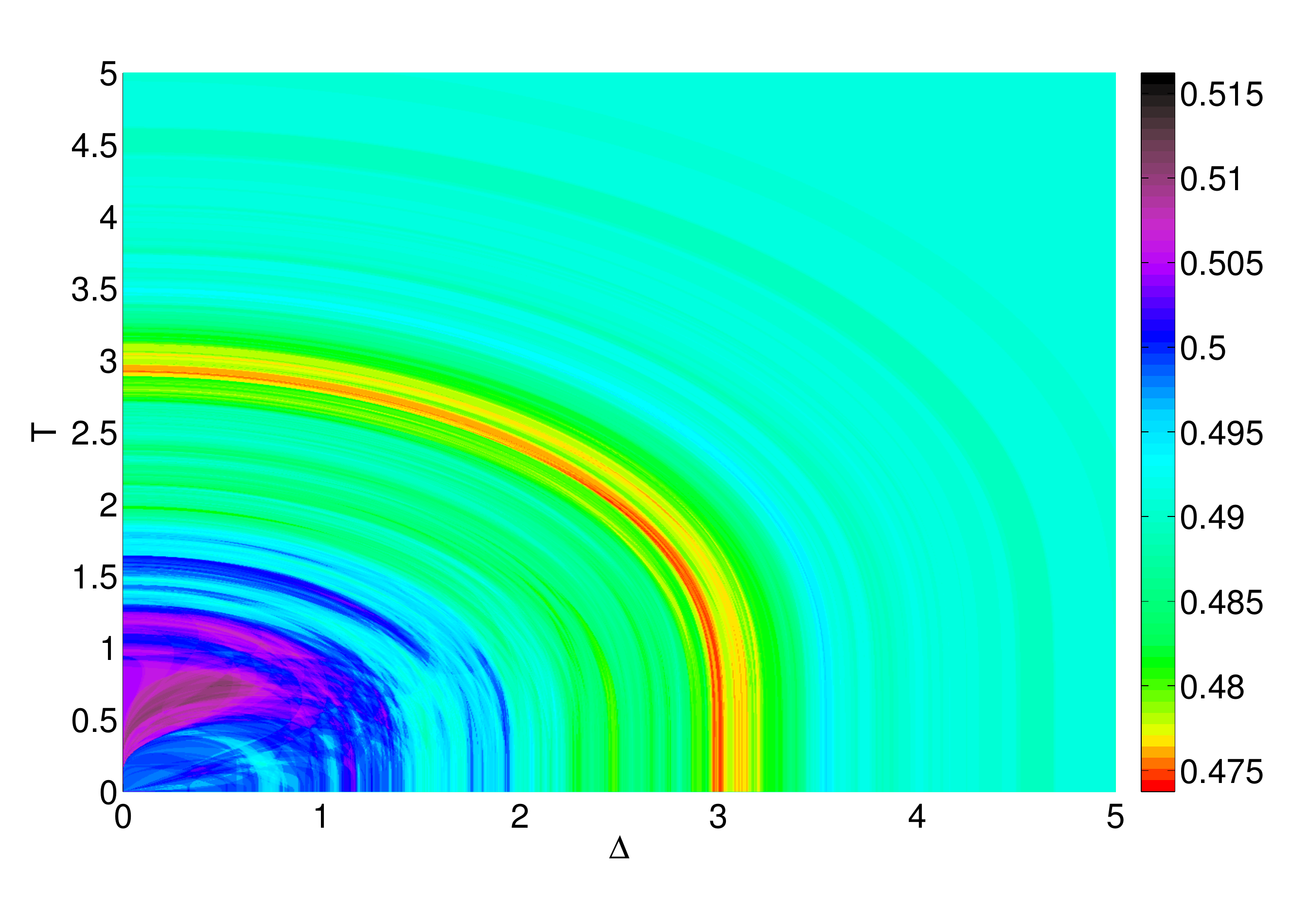}
\par\end{centering}

Figure S2: Colour plots indicating the fraction of bits with spin-sign
transitions which disagree with finite temperature and transverse
field states for the effective superspin Hamiltonian for $\alpha=1$
and $\alpha_{s}=0.1$.
\end{figure}

\section{Full unit cell with $2000\mu s$ annealing time}

While we are able to demonstrate equilibration with a truncated unit
cell as illustrated in Fig. 5 of the main document, we were not able
to see the same kind of behavior for superspins based on the full
chimera unit cell. Fig. S3 illustrates that even if the energy scale
$\alpha$ is reduced to $0.15$ and the annealing time increased to
$2000\,\mu s$, the data still show remnants of the transverse field.

\begin{figure}
\begin{centering}
\includegraphics[width=7cm]{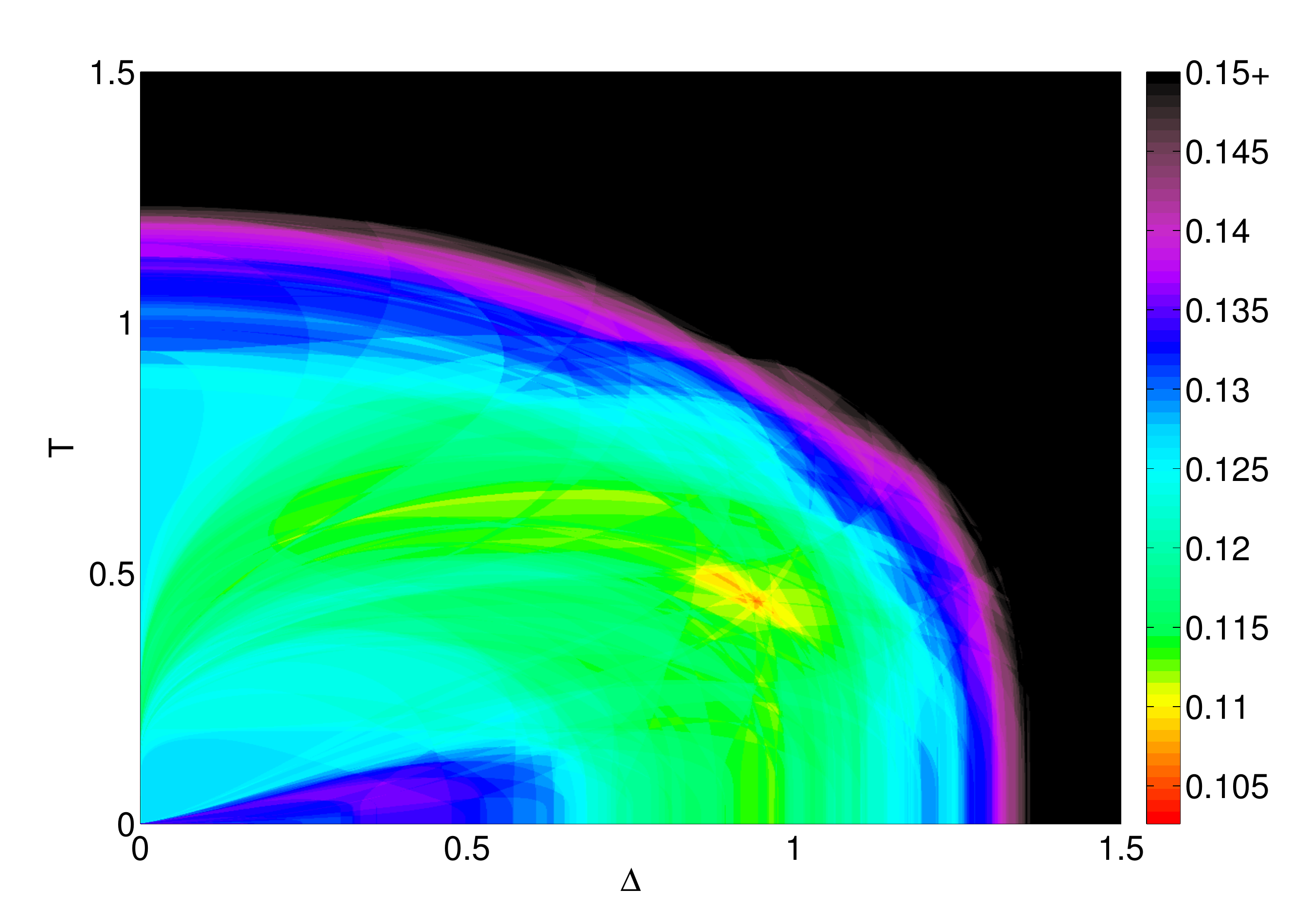}
\par\end{centering}

Figure S2: Colour plots indicating the fraction of bits with spin
sign transitions which disagree with finite temperature and transverse
field states for the effective superspin Hamiltonian for $\alpha=0.15$
, $\alpha_{s}=1$ and a run time of $2000\,\mu s$.
\end{figure}

\end{document}